\documentstyle[12pt]{article}

\begin{document}

\begin{center}

{\Large{\bf Probabilistic Approach to Pattern Selection} \\ [5mm]

V.I. Yukalov} \\ [2mm]

{\it
Bogolubov Laboratory of Theoretical Physics \\
Joint Institute for Nuclear Research, Dubna 141980, Russia \\
and\\
Instituto de Fisica de S\~ao Carlos, Universidade de S\~ao Paulo\\
Caixa Postal 369, S\~ao Carlos, S\~ao Paulo 13560-970, Brazil}

\end{center}

\vskip 2cm

\begin{abstract}

The problem of pattern selection arises when the evolution equations have
many solutions, whereas observed patterns constitute a much more restricted
set. An approach is advanced for treating the problem of pattern selection
by defining the probability distribution of patterns. Then the most probable
pattern naturally corresponds to the largest probability weight. This
approach provides the ordering principle for the multiplicity of solutions
explaining why some of them are more preferable than other. The approach is
applied to solving the problem of turbulent photon filamentation in resonant
media.

\end{abstract}

\vskip 1cm

{\bf PACS codes:} 02.30.Jr, 42.65.Sf

\vskip 5mm

{\bf Key words:} Problem of pattern selection, Turbulent photon 
filamentation

\newpage

\section{Introduction}

Nonequilibrium phenomena in real systems are usually described by sets
of nonlinear differential or integro-differential equations in partial
derivatives. The solutions to such equations are in many cases nonuniform
in space exhibiting the formation of various spatial structures. It often
happens that a given set of equations possesses several solutions corresponding
to different spatial patterns. And, moreover, it is often impossible to
distinguish between these solutions being based on stability analysis since
all these solutions can be stable. If an ensemble of solutions can be
parametrized by a multiparameter $\beta$ from a manifold ${\cal B}$, so that
all solutions corresponding to any $\beta\in{\cal B}$ are stable, then the
manifold ${\cal B}$ is called the {\it stability balloon} [1].

In the case of such a nonuniqueness of solutions one may wander which of the
multiplicity of allowed states would actually be found for a given experimental
protocol. This is how the {\it problem of pattern selection} arises, when the
considered equations have many solutions for given external conditions,
whereas observed patterns may constitute a more restricted set. And there no
doubts are numerous systems and phenomena in real life with such a complicated
behaviour [1].

Not only the solutions to equations can be nonunique but real phenomena may
also display a variety of patterns for the same given conditions. This means
that the multiplicity of solutions is not just an artifact but a feature
expressing the intrinsic complexity of natural phenomena. If real patterns
show multiplicity, is there any ordering between them, such that one solution
is preferred over the other? In equilibrium thermodynamics this is a familiar
concept, with the free energy providing the ordering principle. But in
nonequilibrium systems, there is no such a general organizing principle to
apply [1]. For some particular cases several heuristic recipes have been
suggested. For example, one state may be preferred over the other if it has a
larger basin of attraction for typical initial conditions. Or one may think
that the fastest growing mode dominates the evolution. Or selection is made
assuming that the pattern with the fastest spatial decay rate is preferable.
The limits of such heuristic arguments are discussed in detail in Cross and
Hehenberg [1] where one can find extensive citations to literature on the
problem.

In the present paper a general approach is developed for treating the problem
of pattern selection. The approach, which is formulated in Sec. 2, is based
on defining the probability distribution of patterns. In Sections 4 to 5,
this approach is applied to the problem of turbulent photon filamentation.
An introduction to the latter problem is given in Section 3. The obtained
results are in very good agreement with experiment. The final Section 7
contains conclusions.

\section{Probability Distribution of Patterns}

Let us consider a system of evolution equations, which displays the
multiplicity of solutions describing different spatial structures. Assume
that these solutions can be parametrized by a multiparameter $\beta$. All
admissible values of $\beta$ form a manifold ${\cal B}=\{\beta\}$. In many
cases, differential equations in partial derivatives can be reduced to a
$d$-dimensional system of ordinary differential equations, with a dimension
$d$ that may equal infinity [2]. For the simplicity of notation, we shall
keep in mind this possibility of working with a $d$-dimensional dynamical
system. A generalization of this case will be given at the end of the present
Section.

The state of a $d$-dimensional dynamical system is the set
\begin{equation}
\label{1}
y(t) =\{ y_i(t)=y_i(\beta,t)|\; i=1,2,\ldots,d\}
\end{equation}
of solutions to the system of differential equations, which can always be
presented in the normal form
\begin{equation}
\label{2}
\frac{d}{dt}\; y(t) = v(y,t) \; ,
\end{equation}
where $v=\{ v_i|\; i=1,2,\ldots,d\}$ is a velocity field. By assumption, the
multiplicity of solutions is parametrized by a multiparameter $\beta$, but
in intermediate calculations we shall not, for the compactness of notation,
label functions with $\beta$, restoring this dependence in final formulas.

Our aim is to classify the states (1) labelled by $\beta$ by defining a
probability measure on the manifold ${\cal B}$. To introduce the probability
distribution $p(\beta,t)$ of patterns labelled by a multiparameter $\beta$,
we may resort to the ideas of statistical mechanics [3], where a probability
$p$ can be connected with entropy $S$ by the relation $p\sim e^{-S}$. In the
nonequilibrium case, the entropy is a function of time, $S(t)$. Since it is
not the entropy itself but rather its change that is measurable, it is
convenient to count entropy from its initial value $S(0)$, thus, considering
the entropy variation
\begin{equation}
\label{3}
\Delta S(t) = S(t) - S(0) \; ,
\end{equation}
which is a kind of relative entropy [4]. The entropy may be defined as the
logarithm of an elementary phase volume [3], which for the nonequilibrium
case can be expressed as
\begin{equation}
\label{4}
S(t) =\ln| \delta\Gamma(t)| \; ,
\end{equation}
with the elementary phase volume
\begin{equation}
\label{5}
\delta\Gamma(t) \equiv \prod_i \delta y_i(t) \; .
\end{equation}
Hence the entropy variation (3) is
\begin{equation}
\label{6}
\Delta S(t) =\ln\left |\frac{\delta\Gamma(t)}{\delta\Gamma(0)}\right |\; .
\end{equation}
The probability distribution $p\sim e^{-\Delta S}$, being normalized
with respect to $\beta$, takes the form
\begin{equation}
\label{7}
p(\beta,t) =\;\frac{e^{-\Delta S(\beta,t)}}{Z(t)} \; , \qquad
Z(t) \equiv \int\; e^{-\Delta S(\beta,t)}\; d\beta \; ,
\end{equation}
where the integration over $\beta$ runs through the manifold ${\cal B}$. If
the latter manifold is not continuous, the integration is to be replaced by
summation.

The elementary phase volume (5) can be presented as
\begin{equation}
\label{8}
\delta\Gamma(t) = \prod_i\sum_j M_{ij}(t) \delta y_j(0) \; ,
\end{equation}
where $M_{ij}(t)$ are the elements of the multiplier matrix [5] defined
through the variational derivatives
\begin{equation}
\label{9}
M_{ij}(t) \equiv \frac{\delta y_i(t)}{\delta y_j(0)} \; , \qquad
M_{ij} (0) =\delta_{ij} \; .
\end{equation}
Then the entropy variation (6) writes
\begin{equation}
\label{10}
\Delta S(t) = \sum_i \ln| M_{ii}(t)| \; .
\end{equation}
The multiplier matrix $\hat M(t) = [M_{ij}(t)]$ satisfies the equation
\begin{equation}
\label{11}
\frac{d}{dt}\;\hat M(t) = \hat J(y,t) \hat M(t) \; ,
\end{equation}
which follows from the variation of the evolution equation (2) and where
$\hat J(y,t) =[J_{ij}(y,t)]$ is the Jacobian matrix with the elements
\begin{equation}
\label{12}
J_{ij}(y,t) = \frac{\delta v_i(y,t)}{\delta y_j(t)} \; .
\end{equation}
The initial condition for Eq. (11), according to the definition (9), is
$\hat M(0)=[\delta_{ij}]$. Combining Eqs. (7) and (10), we get the {\it
probability distribution of patterns},
\begin{equation}
\label{13}
p(\beta,t) =\; \frac{1}{Z(t)}\; \prod_i\; \frac{1}{|M_{ii}(\beta,t)|} \; ,
\end{equation}
with the normalization factor
$$
Z(t) = \int \prod_i \; \frac{1}{|M_{ii}(\beta,t)|}\;d\beta\; .
$$
This is the probability distribution of solutions classified with a
multiparameter $\beta$. Since each solution, by definition, represents a
particular pattern, expression (13) is the probability distribution of
patterns. This expression naturally connects the notion of probability
and the notion of stability. Really, the multipliers are smaller by modulus
for more stable solutions and, consequently, for more probable patterns.

To calculate the probability distribution of patterns (13), we need to know
the multipliers (9) which are defined by the equation (11). Using the latter
equation, the pattern distribution (13) can be transformed as follows.
Introduce the matrix $\hat L(t) =[L_{ij}(t)]$ with the elements
\begin{equation}
\label{14}
L_{ij}(t) \equiv \ln| M_{ij}(t)| \; .
\end{equation}
Then the entropy variation (10) writes
\begin{equation}
\label{15}
\Delta S(t) ={\rm Tr}\;\hat L(t) \; .
\end{equation}
The trace of a matrix does not depend on the matrix representation. Hence,
we may perform intermediate calculations using one particular representation,
returning at the end to the form independent of a representation. To this
end, let us consider a representation when the multiplier matrix is diagonal.
Because of Eq. (11) with the initial condition $M_{ij}(0)=\delta_{ij}$, the
multiplier matrix is diagonal if an only if the Jacobian matrix is diagonal
too. In this case, the evolution equation (11) yields
$$
M_{ii}(t) =\exp\left\{ \int_0^t \; J_{ii}(y(t'),t')\; dt'\right\} \; .
$$
From Eq. (14) we have
$$
L_{ii}(t) = \int_0^t\; {\rm Re}\; J_{ii}(y(t'),t')\; dt'\; .
$$
Introducing the notation
\begin{equation}
\label{16}
K(t) \equiv \sum_i\; {\rm Re}\; J_{ii}(y,t) \; ,
\end{equation}
we get
$$
{\rm Tr}\; \hat L(t) = \int_0^t\; K(t')\; dt'\; .
$$
Without the loss of generality, we may assume that the state (1) consists
of real functions since any complex function can always be treated as a pair
of real functions. Hence the velocity field in Eq. (2) can also be considered
as real. Then the eigenvalues of the Jacobian matrix (12) are either real or,
if complex, come in complex conjugate pairs. Therefore
$$
\sum_i\; {\rm Re}\; J_{ii}(y,t) = \sum_i\; J_{ii}(y,t) ={\rm Tr}\;
\hat J(y,t) \; .
$$
Thus, Eq. (16) can be written as
\begin{equation}
\label{17}
K(t) ={\rm Tr}\; \hat J(y,t) \; .
\end{equation}
And for the entropy variation (15), we find
\begin{equation}
\label{18}
\Delta S(t) = \int_0^t\; K(t')\; dt'\; .
\end{equation}

Expression (17) in dynamical theory is called the {\it contraction rate}
[6]. All consideration given above can be straightforwardly generalized to
the case when the state (1) consists of functions $y_i(x,t)$, where $x$ is
the set of space variables. This results not more than in a slight complication
of notations. Then everywhere the variable $x$ appears together with the
index $i$ as an additional continuous index, and the sums over $i$ are to be
complimented by the integrals over $x$. The multiplier matrix (9) becomes a
matrix with respect to indices $i,\;j$ as well as with respect to $x,\;x'$,
$$
M_{ij}(x,x',t) \equiv \frac{\delta y_i(x,t)}{\delta y_j(x',0)}\; ,
$$
with the initial condition
$$
M_{ij}(x,x',0) =\delta_{ij}\delta(x-x') \; .
$$
Similarly, the Jacobian matrix (12) becomes a matrix with the elements
$$
J_{ij}(x,x',y,t) \equiv \; \frac{\delta v_i(x,y,t)}{\delta y_j(x',t)}\; .
$$
Employing the matrix notation with respect to continuous variables [5], we
may repeat the same steps as above in deriving the probability distribution
(13). As is mentioned, the sum over $x$ is to be understood as the
corresponding integral. And the product over a continuous variable can be
defined [7] as
$$
\prod_x \; f(x) \equiv \exp \int \ln f(x)\; dx \; .
$$
As a result, the pattern distribution (13) reads
$$
p(\beta,t)=\; \frac{1}{Z(t)}\;\exp\left\{ -\sum_i\;\int\;
\ln|M_{ii}(x,x,\beta,t)|\; dx\right\} \; .
$$
The contraction rate (17) becomes
\begin{equation}
\label{19}
K(t) =\sum_i\; \int\; J_{ii}(x,x,y,t)\; dx \; ,
\end{equation}
where $y=\{ y_i(x,t)\}$. In this way, the contraction rate has always the form
of ${\rm Tr}\hat J(y,t)$, as in Eq. (17), where the trace has to be defined
according to the representation of the Jacobian matrix.

Restoring in the contraction rate the dependence on the parameter $\beta$
labelling different patterns and using the entropy variation (18), we finally
obtain the {\it probability distribution of patterns}
\begin{equation}
\label{20}
p(\beta,t) =\; \frac{1}{Z(t)}\;\exp\left\{ -\int_0^t\;
K(\beta,t')\; dt'\right\} \; ,
\end{equation}
in which the normalization factor is
$$
Z(t) = \int \exp\left\{ -\int_0^t\; K(\beta,t')\; dt'\right\} \;
d\beta\; .
$$

Thus, each solution labelled by $\beta$ is equipped with the probability
weight (20). Consequently, that pattern is preferred over the other which
has a higher probability weight. This is equivalent, because of the form
(20), to saying that one pattern is preferable over others if its {\it
local contraction}
$$
\Lambda(\beta,t) \equiv \; \frac{1}{t} \;
\int_0^t\; K(\beta,t')\; dt'
$$
is minimal with respect to $\beta$. The latter provides the ordering
principle for classifying solutions and related patterns. The local 
contraction plays for nonequilibrium dynamical systems the role analogous 
to that of the free energy for equilibrium statistical systems.

\section{Turbulent Photon Filamentation}

To illustrate the probabilistic approach to pattern selection, developed in
the previous section, let us consider spatial structures appearing in resonant
samples when increasing the Fresnel number. In active media interacting with
electromagnetic field, there appears a variety of spatiotemporal structures.
For example, electric field in laser cavities can exhibit spatial states, such
as solitons and vortices, which bear a close analogy with similar structures
in liquids [8,9]. In general, there is a  direct correspondence between the
Maxwell-Bloch equations for slowly varying field amplitudes and hydrodynamic
equations for compressible viscous liquid [1,10]. The Fresnel number for
optical systems plays the same role as the Reynolds number for fluids. In the
same way as when increasing the Reynolds number, the fluid becomes turbulent,
the field dynamics exhibits chaotic behaviour when increasing the Fresnel
number. Similarly to hydrodynamics, optical systems display spatial
multi-stability with several coexisting distinct stable states for the same
values of parameters [1,11]. Spatiotemporal chaos in optical systems is
characterized by the same fast decay of correlation functions as it occurs
for turbulent fluids. This is why one calls such chaotic optical phenomena
{\it optical turbulence} [1,10,11].

In active optical systems, having the standard cylindrical shape, the
transition from regular behaviour to spatiotemporal chaos occurs with
the increasing Fresnel number
$$
F\equiv \; \frac{\pi R^2}{\lambda L} \; ,
$$
where $R$ is the cylinder radius, $L$ is the characteristic length, and
$\lambda$ is the radiation wavelength. Physical processes accompanying the
route from a regular regime to chaotic one are similar in different active
optical media.

At small Fresnel numbers $F\ll 1$, there exists the sole transverse central
mode practically uniformly filling the medium. When the Fresnel number is
around $F\sim1$, the cavity can house several transverse modes seen as a
regular arrangement of bright spots in the transverse cross-section. These
regular spatial structures emerge from an initially homogeneous state with
a break of space-translational symmetry. They are regular in space forming
ordered geometric arrays, such a polygons, and they are regular in time
being either stationary or periodically oscillating. These transverse
structures are imposed by the cavity geometry and correspond to the
empty-cavity Gauss-Laguerre modes. Such regular structures are well
understood theoretically, their description being based on field expansions
over the modal Gauss-Laguerre functions prescribed by the cylindrical
geometry, and the theory being in reasonable agreement with experiments
for lasers, e.g. for CO$_2$ and Na$_2$ lasers [11--17], and for active
nonlinear media, as the photorefractive Bi$_{12}$SiO$_{20}$ crystal pumped
by a laser [18--20]. Similar structures also arise in many passive nonlinear
media, such as a Kerr medium [20]. For Fresnel numbers up to $F\approx 5$,
the number of bright spots is proportional to $F^2$. In the longitudinal
cross-section this corresponds to the existence of bright filaments whose
number follows the $F^2$ law as $F$ increases.

Around $F\approx 10$ there occurs a principal change of properties, from the
existence of regular structures to a turbulent-type state [21--23], with the
intermittent behaviour in the region $5<F<15$. This happens in lasers [21--23]
as well as in photorefractive crystals [18--20]. For Fresnel numbers $F>15$,
the arising spatial structures are very different from those associated with
the empty-cavity modes. The modal expansion is no longer relevant and the
boundary conditions have no importance. The medium looks as consisting of
a large number of parallel independently flashing filaments, whose number is
proportional to $F$, contrary to the case of small Fresnel numbers with the
number of filaments proportional to $F^2$. The filaments are chaotically
distributed in space, are not correlated with each other, and are aperiodically
flashing in time. Such a spatio-temporal chaotic behaviour is characteristic
of hydrodynamic turbulence, this is why the similar phenomenon in optics is
commonly called the {\it optical turbulence} [1,10,11,23]. Contrary to the
regime of small $F$, when the regular spatial structures are prescribed by
the geometry and boundary conditions imposing their symmetry constraints, the
turbulent optical filamentation is strictly self-organized, with its
organization emerging from intrinsic properties of the medium [20,24]. This
kind of optical turbulence has been observed in both photorefractive crystals
[18--20] and lasers [21--23,25--30]. Especially accurate and thorough
experimental studies for CO$_2$ and Dye lasers have been accomplished in
Refs. [25--30]. The independence of turbulent optical filamentation of
boundary conditions and, hence, its purely self-organized nature are confirmed
by the observation of this effect in the resonatorless discharge-tube
superluminescent samples, such as lasers on Ne, Tl, Pb, N$_2$, and N$_2^+$
vapors [31--35]. Since the optical turbulence is characterized by the
formation of bright filaments with a high density of photons, this phenomenon
may be named the {\it turbulent photon filamentation}.

Let us summarize the main features characterizing the regular and turbulent
regimes in active optical media. At low Fresnel numbers $F<10$, the regular
regime occurs whose typical characteristics are: (i) Bright filaments are
regularly arranged in space forming an ordered structure seen in the transverse
cross-section as a polygon made of bright spots. (ii) The spatially ordered
structure is either stationary or periodically oscillating in time. (iii)
The number of bright filaments is proportional to $F^2$.

At high Fresnel number $F>10$, the turbulent regime develops whose typical
features are: (i) The bright photon filaments are chaotically distributed in
space. (ii) The filaments aperiodically flash in time. (iii) The number of
these uncorrelated filaments is proportional to $F$.

While for the low-Fresnel-number regime, with regular optical structures, a
good theoretical understanding was achieved [11--20], for the high-Fresnel-number
regime of the turbulent photon filamentation, there has been no persuasive
theory developed. This problem was addresses in Refs. [36--39], where the
consideration was based on a simple model, only the stationary case was
analysed, and the minimum-energy arguments were employed. This model
consideration showed that photon filaments can really be formed in resonant
media due to an effective interaction between atoms through photon exchange.
The estimates for the filament radius turned out to be in good agreement
with experiment. However, the empirical approach of Refs. [36--39] cannot
be accepted as completely satisfactory. This is mainly because of the two
principal points. One is that a stationary case was addressed, while
turbulence is rather a nonstationary phenomenon and must be treated by
time-dependent evolution equations. The second point is that the
minimal-energy condition was invoked, while there is no such a principle
for nonequilibrium systems.

Here we demonstrate that the problem of turbulent photon filamentation can
be successfully treated by the probabilistic approach to pattern selection
formulated in Section 2. Our consideration is based on realistic evolution
equations for resonant media. Such equations are necessary for calculating
the contraction rate $K$ defining the probability distribution of
patterns (20).

\section{Equations for Resonant Medium}

The system of $N$ resonant atoms interacting with electromagnetic field is
described by the Hamiltonian
\begin{equation}
\label{21}
\hat H =\hat H_a +\hat H_f +\hat H_{af} \; ,
\end{equation}
consisting of the following terms: The Hamiltonian of two-level resonant
atoms
\begin{equation}
\label{22}
\hat H_a =\frac{1}{2}\; \sum_{i=1}^N \; \omega_0(1+\sigma_i^z)\; ,
\end{equation}
with the transition frequency $\omega_0$, where $\sigma_i^z$ is a Pauli
operator. The radiation-field Hamiltonian
\begin{equation}
\label{23}
\hat H_f =\; \frac{1}{8\pi}\; \int\; \left (\vec E^2 +\vec H^2\right )\;
d\vec r \; ,
\end{equation}
with electric field $\vec E$ and magnetic field $\vec H=\vec\nabla\times
\vec A$, where the vector potential $\vec A$ is assumed to satisfy the
Coulomb gauge calibration $\vec\nabla\cdot\vec A=0$. The atom-field
interaction Hamiltonian
\begin{equation}
\label{24}
\hat H_{af} = -\;\sum_{i=1}^N \left ( \frac{1}{c}\; \vec j_j\cdot\vec A_i +
\vec d_i\cdot\vec E_{0i}\right )
\end{equation}
has the standard dipole form, in which $\vec A_i\equiv\vec A(\vec r_i,t)$;
the transition-current and transition dipole operators are
$$
\vec j_i = i\omega_0\left ( \vec d\sigma_i^+ -\vec d^*\sigma_i^-\right )\; ,
\qquad \vec d_i=\vec d\sigma_i^+ +\vec d^*\sigma_i^- \; ,
$$
where $\vec d$ is the transition dipole and $\sigma_i^\pm$ are the rising or
lowering operators, respectively; and $\vec E_{0i}$ is a seed field [40].

To consider nonuniform systems with arising spatial structures, it is
convenient to invoke the space-time representation of evolution equations,
whose general description is given in the book [40] and a detailed analysis
in Refs. [41,42]. The equations are written for the average quantities
\begin{equation}
\label{25}
u(\vec r,t)\equiv\; < \sigma^-(\vec r,t)>\; , \qquad
s(\vec r,t)\equiv\; <\sigma^z(\vec r,t)>\; ,
\end{equation}
using the standard semiclassical and Born approximations. For the compactness
of presentation, let us introduce the notation
\begin{equation}
\label{26}
f(\vec r,t)\equiv f_0(\vec r,t) + f_{rad}(\vec r,t)
\end{equation}
for an effective field acting on an atom. This field consists of the term
\begin{equation}
\label{27}
f_0(\vec r,t) \equiv \; -i\vec d\cdot\vec E_0(\vec r,t) \; ,
\end{equation}
due to the cavity seed field, and of the term
\begin{equation}
\label{28}
f_{rad}(\vec r,t) =\; -\;\frac{3}{4}\; i\gamma\rho\; \int\; \left [
\varphi(\vec r - \vec r\;')\; u(\vec r\;',t) -\vec{e_d}^2\varphi^*(\vec r-
\vec r\; ')\; u^* (\vec r\; ',t)\right ]\; d\vec r\; '
\end{equation}
corresponding to the interaction of atoms through the common radiation field,
where $\rho$ is the density of atoms, $\vec d\equiv d_0\vec e_d,\; d_0\equiv
|\vec d|$, and
$$
\varphi(\vec r) \equiv\; \frac{\exp(ik_0\;|\vec r|)}{k_0\;|\vec r|} \; ,
\qquad k_0 \equiv\; \frac{\omega_0}{c}\; , \qquad \gamma\equiv\;
\frac{4}{3}\; k_0^3d_0^2\; .
$$
The seed field
\begin{equation}
\label{29}
\vec E_0(\vec r,t) =\vec E_1\; e^{i(kz-\omega t)} + \vec E_1^* \;
e^{-i(kz-\omega t)}
\end{equation}
selects a longitudinal mode propagating along the axis $z$ which is the axis
of a cylindrical sample. The resulting equations are
$$
\frac{du}{dt} =\; -(i\omega_0 +\gamma_2)u + sf\; ,
$$
\begin{equation}
\label{30}
\frac{ds}{dt} =\; -2(u^* f + f^* u) -\gamma_1 (s-\zeta) \; ,
\end{equation}
$$
\frac{d|u|^2}{dt} =\; -2\gamma_2|u|^2 + s(u^* f + f^* u) \; ,
$$
where $\gamma_1$ and $\gamma_2$ are the level and line widths, respectively,
and $\zeta$ is the pumping parameter defined by the intensity of nonresonant
lamp pumping.

The sample has cylindrical shape typical of lasers. The radiation wavelength
$\lambda$, cylinder radius $R$, and its length $L$ are related by the
inequalities
\begin{equation}
\label{31}
\frac{\lambda}{R}\ll 1\; , \qquad \frac{R}{L}\ll 1\; .
\end{equation}
We consider the quasiresonance situation when
\begin{equation}
\label{32}
\frac{|\Delta|}{\omega_0}\ll 1 \; , \qquad \Delta =\omega-\omega_0 \; .
\end{equation}
As usual, the relaxation parameters are assumed to be small, so that
\begin{equation}
\label{33}
\frac{\gamma_1}{\omega_0}\ll 1 \; , \qquad \frac{\gamma_2}{\omega_0}\ll 1 \; .
\end{equation}

The solutions to Eqs. (30) are, in general, nonuniform. This is because atoms
interact with each other thorough the radiation field (28) containing the
function $\varphi(\vec r)$ fastly oscillating in space and diminishing with
increasing $|\vec r|$. Hence the atoms, situated far from each other, do not
interact and radiate independently, although in the longitudinal direction
they may be correlated due to the cylindrical geometry of the sample and the
propagation of radiated field along the axis $z$. This suggests to look for a
solution of Eqs. (30) in the form of a bunch of filaments aligned along the
cylinder axis. Let us imagine that the radiation field inside the sample is
stratified into $N_f$ photon filaments stretched along the axis $z$. These
filaments can have different radii $r_n$, with $n=1,2,\ldots,N_f$. The
radiation is mainly concentrated inside the bright filaments, fading away
outside them, so that at the radial distance $b_n$ the corresponding
solutions are an order of magnitude smaller than at the axis of a filament.
The relation between the characteristic lengths $b_n$ and $r_n$ can be written
down if the profile of solutions inside a filament is known. If the filament
profile can be approximated by the normal law $\exp(-r^2/2r_n^2)$, with the
filament radius $r_n$ being the standard deviation, then
\begin{equation}
\label{34}
b_n=\sqrt{2\ln 10}\; r_n \; .
\end{equation}
The filaments do not interact with each other, because of which their
locations in the radial cross-section are random. This means that the filament
axes are located at random points $\{ x_n,y_n\}$, where $n=1,2,\ldots,N_f$.

The above discussion explains why it is reasonable to try to find the solutions
of Eqs. (30) in the form of expansions
\begin{equation}
\label{35}
u(\vec r,t) =\sum_{n=1}^{N_f}\; u_n(\vec r,t)\Theta_n(x,y)\; e^{ikz} \; ,
\qquad
s(\vec r,t) =\sum_{n=1}^{N_f}\; s_n(\vec r,t)\Theta_n(x,y)
\end{equation}
over filaments, where
$$
\Theta_n(x,y) \equiv \Theta\left ( b_n^2 -(x-x_n)^2 + (y-y_n)^2\right )
$$
is a unit-step function. Substituting the presentation (35) in Eqs. (30), we
obtain a system of equations for $u_n,\; s_n$, and $|u_n|^2$. These equations
can be simplified if we are not interested in the detailed internal structure
of each filament but rather wish to find out their characteristic sizes.
Then we may employ the mean-field approximation for the averages
$$
u(t) \equiv\; \frac{1}{V_n}\; \int_{V_n}\; u_n(\vec r,t)\; d\vec r\; , \qquad
s(t) \equiv\; \frac{1}{V_n}\; \int_{V_n}\; s_n(\vec r,t)\; d\vec r\; ,
$$
\begin{equation}
\label{36}
|u(t)|^2 \equiv \; \frac{1}{V_n}\; \int_{V_n}\; |u_n(\vec r,t)|^2\;
d\vec r\; ,
\end{equation}
where the integration is over the volume $V_n\equiv\pi b_n^2L$ and the index
$n$ in the left-hand sides, for short, is dropped.

To present the equations for the functions (36) in a compact form, we introduce
two effective {\it coupling parameters}:
\begin{equation}
\label{37}
g\equiv\; \frac{3\gamma\rho}{4\gamma_2V_n} \; \int_{V_n}\;
\frac{\sin[k_0|\vec r-\vec r\;'| -k(z-z')]}{k_0|\vec r-\vec r\;'|}\;
d\vec r\; d\vec r\;'
\end{equation}
and
\begin{equation}
\label{38}
g'\equiv\; \frac{3\gamma\rho}{4\gamma_2V_n} \; \int_{V_n}\;
\frac{\cos[k_0|\vec r-\vec r\;'| -k(z-z')]}{k_0|\vec r-\vec r\;'|}\;
d\vec r\; d\vec r\;'\; .
\end{equation}
These parameters enter the definitions of the {\it collective frequency} and
{\it collective width}, respectively,
\begin{equation}
\label{39}
\Omega\equiv\omega_0 + g'\gamma_2 s\; , \qquad
\Gamma\equiv \gamma_2 (1 -gs) \; .
\end{equation}
Then from Eqs. (30), employing the introduced notation, for the functions
(36) we obtain the equations
$$
\frac{du}{dt}=\; -(i\Omega+\Gamma) u - is\vec d\cdot\vec E_1\;
e^{-i\omega t} \; ,
$$
\begin{equation}
\label{40}
\frac{ds}{dt} =\; -4g\gamma_2|u|^2 -\gamma_1(s-\zeta) - 4\;{\rm Im}\left (
u^*\vec d\cdot\vec E_1\; e^{-i\omega t}\right ) \; ,
\end{equation}
$$
\frac{d|u|^2}{dt} =\; -2\Gamma|u|^2 + 2s\; {\rm Im}\left ( u^*\vec d\cdot
\vec E_1\; e^{-i\omega t}\right )\; ,
$$
describing the time evolution of a filament. Such a reduction of initial
equations for a bunch of filaments to the set (40) of equations for each of
the filaments has become possible due to the absence of interactions between
filaments. Different filaments can have different radii whose distribution is
yet to be defined.

It is worth emphasizing that the presentation of solutions as the expansions
(35) over filaments is rather general and includes as well the case when
there are no separate filaments at all. Really, if it turns out that the
most probable filament radius is close to the radius of the sample, $R$,
this would actually mean that there are no several filaments, but the whole
volume of the sample is filled by one filament.

\section{Characteristics of Arising Filaments}

To analyze Eqs. (40), we employ the scale separation approach [43,44] which
is a generalization of the averaging technique [45] to the case of
nonequilibrium statistical systems. To this end, we notice first of all that
the functional variables in Eqs. (40) can be classified onto fast and slow
in time. Due to the existence of the small parameters (33), the variable
$u$ is fast as compared to the slow variables $s$ and $|u|^2$. The slow
variables play the role of quasi-invariants for the fast function $u$. With
$s$ being a quasi-invariant, the solution for the fast variable $u$,
described by the first of Eqs. (40), writes
\begin{equation}
\label{41}
u(t) = u_0\; e^{-(i\Omega+\Gamma)t} + \;
\frac{s\vec d\cdot\vec E_1}{\omega-\Omega+i\Gamma}\; \left [
e^{-i\omega t} - e^{-(i\Omega+\Gamma)t}\right ] \; ,
\end{equation}
where $u_0=u(0)$. This solution is to be substituted in the second and third
of Eqs. (40) whose right-hand sides are to be averaged over time explicitly
entering the fastly oscillating functions. In this way, we meet the quantity
\begin{equation}
\label{42}
\alpha\equiv\; \frac{{\rm Im}}{s\Gamma}\; \lim_{\tau\rightarrow\infty}\;
\frac{1}{\tau}\; \int_0^\tau\; u^*(t)\;\vec d\cdot\vec E_1\;
e^{-i\omega t}\; dt
\end{equation}
describing the influence of the seed field on atoms. Taking into account
Eq. (41), we have
\begin{equation}
\label{43}
\alpha=\; \frac{|\vec d\cdot\vec E_1|^2}{(\omega-\Omega)^2+\Gamma^2}\; .
\end{equation}
Recall that the seed field is, by definition, a weak cavity field whose
role is to fix the axis of the cylindrical sample. This field does not pump
the energy into the system which is standardly done by nonresonant lamps and
which is described by the pumping parameter $\zeta$ in the evolution equations
(30) or (40). The seed field selects a longitudinal mode imposing no
restrictions on transverse modes. The selection of the latter has to be done
by the internal properties of the evolution equations. The weakness of the
seed field is to be understood in the sense of the smallness of the parameter
(43),
\begin{equation}
\label{44}
\alpha\ll 1\; .
\end{equation}

In analysing further the evolution equations (40), it is convenient to make
the transformation of the functional variable $|u|^2$ to
\begin{equation}
\label{45}
w\equiv |u|^2 -\alpha s^2 \; .
\end{equation}
Finally, following the scale separation approach [43,44], from the last two
of Eqs. (40) we obtain the equations
\begin{equation}
\label{46}
\frac{ds}{dt}= -4g\gamma_2 w -\gamma_1(s-\zeta) \; ,  \qquad
\frac{dw}{dt}= -2\gamma_2 (1-gs) w 
\end{equation}
for the slow variables. These equations can be written in the form (2),
$$
\frac{ds}{dt} = v_1\; , \qquad \frac{dw}{dt} =v_2 \; ,
$$
with the velocity field
$$
v_1= -4g\gamma_2 w -\gamma_1(s-\zeta) \; , \qquad
v_2= -2\gamma_2(1-gs) w \; .
$$
For the Jacobian matrix (12), we now have
$$
J_{11}=\frac{\partial v_1}{\partial s}= -\gamma_1 \; , \qquad
J_{12}=\frac{\partial v_1}{\partial w}= -4g\gamma_2 \; ,
$$
$$
J_{21}=\frac{\partial v_2}{\partial s}= -2\gamma_2 w \; , \qquad 
J_{22}=\frac{\partial v_2}{\partial w}= -2\gamma_2 (1-gs) \; .
$$
From here, we find the contraction rate (17),
\begin{equation}
\label{47}
K ={\rm Tr}\; \hat J =  -\gamma_1 -2\gamma_2(1-gs) \; .
\end{equation}
The value of Eq. (47) depends on the characteristics of filaments through
the coupling parameter (37) containing in its definition the volume
$V_n=\pi b_n^2L$, which is the volume of a cylinder enveloping a filament.
The radius of the enveloping cylinder, $b_n$, is related to the filament
radius, $r_n$, by relation (34). For what follows, it is convenient to
introduce the dimensionless quantity
\begin{equation}
\label{48}
\beta\equiv\;\frac{kb_n^2}{2L} =\; \frac{\pi b_n^2}{\lambda L} \; .
\end{equation}
Thus, the contraction rate (47) is a function $K=K(\beta,t)$
of the variable (48) and also a function of time entering through $s=s(t)$.
For different filaments, the variable $\beta$ can take different values
from the interval
\begin{equation}
\label{49}
0 < \beta\leq F \; ,
\end{equation}
where $F$ is the Fresnel number. The distribution of the variable $\beta$ in
the interval (49) is given by the probability distribution (20). The maximum
of the latter defines the most probable $\beta$ and, respectively, the most
probable filament radius.

Let us consider the very beginning of the process, when $t\rightarrow +0$, in
order to understand what are the most probable filaments to be formed. At
this initial stage, the probability distribution (20) can be written as
$$
p(\beta,t) \simeq \; \frac{1}{Z(t)}\; \exp\left\{ -K(\beta,0)\; t
\right\} \; .
$$
Hence the maximum of this distribution corresponds to the minimal
contraction rate. The latter, according to Eq. (47), depends on $\beta$
through $g=g(\beta)$. We consider the standard laser-type setup, when the
atoms at the initial time are not inverted, i.e. $s_0<0$, and the medium
is pumped with a long nonresonant pulse whose action is described by the
pumping parameter $\zeta>0$ in the evolution equations (30), (40), and
(46). In this case, the minimum of the contraction rate $K(\beta,0)$
is equivalent to the maximum of the coupling $g(\beta)$, which is defined
by the equations
\begin{equation}
\label{50}
\frac{dg}{d\beta} = 0\; , \qquad
\frac{d^2g}{d\beta^2} < 0 \; .
\end{equation}
To solve these equations, we need to analyse the integral (37) giving
$g=g(\beta)$.

The coupling $g(\beta)$ defined by the integral (37) can be presented in the
form
\begin{equation}
\label{51}
g(\beta) =\;\frac{3\pi\gamma\rho L}{4\gamma_2k^2}\; \left [ \pi\beta -
\int_0^{2\beta} \; {\rm Si}(x)\; dx\right ] \; .
\end{equation}
The procedure of reducing Eq. (37) to the form (51) is described in the
Appendix. The extrema of $g(\beta)$ are given by the equation
\begin{equation}
\label{52}
{\rm Si}(2\beta) =\pi \; .
\end{equation}
From several solutions of Eq. (52) we have to choose the absolute maximum
of $g(\beta)$. The latter occurs at $\beta=0.96$, which, according to
Eq. (48), gives $b_n=0.55\sqrt{\lambda L}$. Using the relation (34), we
find the most probable radius of a filament
\begin{equation}
\label{53}
r_f=0.26\sqrt{\lambda L} \; .
\end{equation}
In the typical experiments [21--23,25--35] observing the turbulent photon
filamentation, the excitation was achieved by means of the quasistationary
nonresonant pumping. Such a pumping can be treated as a quasistationary
process if its duration is much longer than the characteristic time
$2\pi/\omega$ of fast oscillations, which is always the case. The
quasistationary pumping can be characterized by an effective pumping parameter
$\zeta$ in the evolution equations. To describe the finite duration time of
the pumping procedure, one can consider $\zeta=\zeta(t)$ as a slow function
of time, slow in the sense of the inequality
$$
\frac{2\pi}{\omega}\;\left | \frac{d\zeta}{dt}\right | \ll 1 \; .
$$
In this way, for the standard laser setup, the filament radius (53) defines
the most probable pattern for the arising bunch of photon filaments.

The most probable number of filaments can be evaluated from the normalization
condition
\begin{equation}
\label{54}
\frac{1}{V}\; \int s(\vec r,t)\; d\vec r =\zeta(t) \; ,
\end{equation}
where the integration runs over the whole volume of the sample, $V=\pi R^2L$.
To approximately calculate the integral (54), let us consider the moment of
time when the filaments have already been formed and the population
difference inside each filament of the radius $r_f$ has reached a value close
to the pumping parameter $\zeta$. Then we may write
$$
\int s(\vec r,t)\; d\vec r\simeq N_fV_f\zeta + (V - N_fV_f) s_{out}\; ,
$$
where $V_f\equiv\pi r_f^2 L$ and $s_{out}$ is a population difference outside
the filaments. In this case, independently of the value $s_{out}$, condition
(54) yields
\begin{equation}
\label{55}
N_f =\left (\frac{R}{r_f}\right )^2 \; .
\end{equation}
For the filaments of the most probable radius (53), equation (55) gives
$$
N_f = 4.71 F\; .
$$
The number of filaments is proportional to the Fresnel number, which is in
complete agreement with all experiments on the turbulent photon filamentation.
The coefficient $4.71$ is also in good agreement with experiments [18--20].

\section{Dynamics of Filament Flashing}

The time evolution of the filaments is described by Eqs. (46). At the initial
stage of the process, when $\gamma_1t\ll 1$, we may omit the term containing
$\gamma_1$. Then the resulting system of nonlinear equations can be solved
exactly yielding the solutions
\begin{equation}
\label{56}
s=-\;\frac{\gamma_0}{g\gamma_2}\; {\rm tanh}\left (\frac{t-t_0}{\tau_0}
\right ) +\;\frac{1}{g} \; , \qquad
w=\;\frac{\gamma_0^2}{4g^2\gamma_2^2}\;{\rm sech}^2\left (
\frac{t-t_0}{\tau_0}\right ) \; ,
\end{equation}
in which $\gamma_0$ is the radiation width, $\tau_0$ is the radiation time,
so that
\begin{equation}
\label{57}
\gamma_0^2 =\Gamma_0^2 + 4g^2\gamma_2^2\left ( |u_0|^2 -\alpha_0 s_0^2
\right ) \; ,
\end{equation}
where $u_0\equiv u(0),\; s_0\equiv s(0),\; \alpha_0\equiv\alpha(0)$, and
\begin{equation}
\label{58}
\Gamma_0^2 \equiv \gamma_2 (1 -gs_0) \; , \qquad \gamma_0\tau_0 \equiv 1;
\end{equation}
the delay time $t_0$ being given by the expression
\begin{equation}
\label{59}
t_0 =\; \frac{\tau_0}{2}\;\ln\left |
\frac{\gamma_0-\Gamma_0}{\gamma_0+\Gamma_0}\right | \; .
\end{equation}
Invoking the transformation (45), we get
\begin{equation}
\label{60}
|u|^2 =\;\frac{\gamma_0^2}{4g^2\gamma_2^2}\;{\rm sech}^2\left (
\frac{t-t_0}{\tau_0}\right ) +\alpha s^2 \; .
\end{equation}
The found solutions depend on the coupling $g$ both directly as well as
through the characteristic parameters $\gamma_0,\; \tau_0$, and $t_0$. This
means that the properties of filaments with different radii are essentially
different, since $g=g(\beta)$.

At the later stage, the term with $\gamma_1$ cannot be dropped. Then Eqs. (46)
cannot be solved exactly. But we may look at the long-time behaviour, when the
system approaches a stationary regime. The latter can be achieved if the
duration of the pumping pulse is longer than $\gamma_1^{-1}$ as well as
$\gamma_2^{-1}$, which is usually the case when pumping a resonant medium.

Before the stationary regime is reached, the oscillation of solutions is
described by the eigenvalues of the Jacobian matrix associated with Eqs. (46).
These eigenvalues are
\begin{equation}
\label{61}
\lambda^\pm =-\;\frac{1}{2}\; \left [ \gamma_1 +\gamma_2(1-gs)\right ]\mp
\left\{ [\gamma_1 -2\gamma_2(1-gs) ]^2 - 32\gamma_2^2 g^2 w
\right \}^{1/2}\; .
\end{equation}
Because of the dependence of $s(t)$ and $w(t)$ on time, Eq. (61) shows that
$\lambda^\pm(t)$ is also time dependent, which means that the oscillation of
solutions cannot be defined by fixed frequencies but occurs in an aperiodic
way.

Al long times, such that $\gamma_1t\gg 1$ and $\gamma_2t\gg 1$, the system
approaches a stationary regime. There are two types of stationary solutions
to Eqs. (46), one of them is given by the fixed point
\begin{equation}
\label{62}
s_1^*=\zeta\; , \qquad w_1^* = 0 ;
\end{equation}
and another, by the fixed point
\begin{equation}
\label{63}
s_2^* =\;\frac{1}{g}\; , \qquad
w_2^*=\; \frac{\gamma_1(g\zeta-1)}{4\gamma_2g^2}\; .
\end{equation}
The stability of the stationary solutions can be determined from the
Lyapunov analysis. To this end, we have to evaluate the Jacobian eigenvalues
(61) at the corresponding fixed points (62) and (63), which results in the
characteristic exponents
\begin{equation}
\label{64}
\lambda_1^+=-\gamma_1\; , \qquad \lambda_1^-=-2\gamma_2(1-g\zeta)
\end{equation}
and, respectively,
\begin{equation}
\label{65}
\lambda_2^\pm =-\;\frac{\gamma_1}{2}\left\{ 1 \pm \left [
1 +8\;\frac{\gamma_2}{\gamma_1}\; ( 1-g\zeta)\right ]^{1/2}\right\} \; .
\end{equation}
The real parts of Eqs. (64) and (65) define the Lyapunov exponents whose
signs characterize the stability of the related fixed points. As far as the
values of the characteristic exponents (64) and (65) depend on the coupling
$g=g(\beta)$ which, according to Eq. (51), essentially depends on the
parameter $\beta$, that is, on the radius of a filament, the stability
properties for the filaments of different radii can be drastically
different. For the parameter $\beta$ in the interval (49), the coupling
(51) varies in a wide diapason. Thus, for small $\beta$, we have
$$
g(\beta) \simeq \; \frac{3\gamma\rho\lambda^2L}{16\gamma_2}\; \beta
\qquad (\beta\ll 1) \; ,
$$
hence $g(\beta)\rightarrow 0$ as $\beta\rightarrow 0$. For large $\beta$,
Eq. (51) gives
$$
g(\beta)\simeq \; \frac{3\gamma L}{16\pi\gamma_2\lambda}\; \rho\lambda^3
\qquad (\beta\gg 1)\; .
$$
In the case of the wavelength $\lambda\gg a$, much larger than the mean
interatomic distance, one has $\rho\lambda^3\gg 1$. Since usually $\gamma\sim
\gamma_2$ and $L\gg\lambda$, the coupling may reach quite large values
$g(\beta)\gg 1$. Such a wide variation of the coupling $g=g(\beta)$ results
in rather different characteristic exponents (64) and (65) for different
filaments and, as a consequence, in essentially different stability properties
of the latter.

All filaments can be separated into three types. One group consists of those
filaments for which
\begin{equation}
\label{66}
g(\beta)\zeta < 1\; .
\end{equation}
Then the fixed point (62) is a stable node, while that (63) is a saddle point.
The filaments with the radii satisfying condition (66) are characterized by
the stationary solutions (62), which shows that these filaments after the time
$T_1\equiv\gamma_1^{-1}$ practically stop radiating coherently.

Increasing $g(\beta)$, one reaches the equality $g(\beta)\zeta=1$, when both
fixed points (62) and (63) merge together becoming neutral. In the interval
\begin{equation}
\label{67}
1 < g(\beta)\zeta\leq 1 +\; \frac{\gamma_1}{8\gamma_2}\; ,
\end{equation}
the fixed point (62) is a saddle point and that (63) is a stable node.
Hence, the filaments with the radii satisfying condition (67) are described
by the stationary solutions (63). Such filaments continue radiating
coherently in the stationary regime, although the level of coherence,
characterized by the value of $w_2^*$, is not high.

The third group of filaments corresponds to large coupling, such that
\begin{equation}
\label{68}
g(\beta)\zeta > 1 +\;\frac{\gamma_1}{8\gamma_2} \; .
\end{equation}
Then the fixed point(62) remains a saddle point while that (63) becomes a
stable focus. The latter means that the approach of the related solutions to
the stationary point (63) is not monotonic but through a series of pulses.
The filaments with the radii satisfying condition (68) radiate by bright
flashes interrupted by the intervals of darkness. This intermittent behaviour,
for finite times, is not periodic, as is seen from the form of Eq. (61), but
as $t\rightarrow\infty$, there appears an asymptotic period. This can be
noticed from the characteristic exponent (65) which, for the case (68),
can be written as
\begin{equation}
\label{69}
\lambda_2^\pm = -\;\frac{\gamma_1}{2}\mp i\omega_\infty \; ,
\end{equation}
where the asymptotic frequency is
\begin{equation}
\label{70}
\omega_\infty =\frac{\gamma_1}{2}\;\left\{8\;\frac{\gamma_2}{\gamma_1}\;
[ g(\beta)\zeta - 1] - 1\right\}^{1/2} \; .
\end{equation}
The value of the latter, because of the dependence on $g(\beta)$, is
essentially different for different filaments.

\section{Conclusion}

An approach is developed for treating the problem of pattern selection. The
approach is based on defining the probability distribution of patterns. This
gives the ordering principle for the multiplicity of admissible solutions to
nonlinear differential equations and, respectively, to the corresponding
space structures. The maximum of the probability distribution naturally
defines the most probable pattern.

This probabilistic approach is applied to the problem of turbulent photon
filamentation in laser media. The most probable filament radius is found,
and the number of filaments is evaluated. The found values are in very good
agreement with experiment. Thus, the filament radii observed in experiments
with dye and CO$_2$ lasers [25--30] are $r_f\approx 0.01$ cm for dye lasers
and $r_f\approx 0.08$ cm for CO$_2$ lasers, which is in perfect agreement
with formula (53). The filament radii measured in experiments with Ne, Tl,
Pb, N$_2$, and N$_2^+$ vapor lasers [31--35] are $r_f\approx 0.01$ cm,
which again coincides with the result of Eq. (53) for the  corresponding
wavelengths and laser lengths. The number of filaments, given by Eq. (55)
is $N_f\approx 10^2-10^3$ for different experiments [31--35]. The found
dependence of the filament radius $r_f\sim F^{-1/2}$ and filament number
$N_f\sim F$ on the Fresnel number is the same as observed experimentally.

The turbulent photon filamentation provides a good example of how a rather
complicated phenomenon can be successfully described by the probabilistic
approach to pattern selection. The formulation of this approach in Section
2 is general, which permits one to apply it for different phenomena where
the problem of pattern selection arises.

\vskip 1cm

{\bf Acknowledgement}

\vskip 3mm

This work is supported by the Bogolubov-Infeld Grant of the State
Agency for Atomic Energy, Poland and by a Grant of the S\~ao Paulo State
Research Foundation, Brazil.

\newpage

{\Large{\bf Appendix. Effective Coupling}}

The effective coupling $g$ is given by the integral (37) over the enveloping
cylinder of radius $b_n$. Assuming that $\lambda\ll b_n$, the integral (37)
can be reduced to the form
$$
g=\;\frac{3\pi\gamma\rho}{2\gamma_2} \; \int_0^{b_n} \; r\; dr \;
\int_{-L/2}^{L/2}\; \frac{\sin(k_0\sqrt{r^2+z^2} - kz)}{k_0\sqrt{r^2+z^2}}\;
dz\; ,
$$
where $r$ is the radial variable. Because of the quasiresonance condition
(32), we have $k_0\approx k$. Then, with the change of the variable
$x\equiv k(\sqrt{r^2+z^2}-z)$, we get
$$
g=\;\frac{3\pi\gamma\rho}{2\gamma_2k}\; \int_0^{b_n}\; r\; dr\;
\int_{kr^2/L}^{kL}\; \frac{\sin x}{x}\; dx\; .
$$
Here, the integration limit $kL$, due to the inequality $\lambda\ll L$, can
be replaced by $\infty$. As a result, we have
$$
g=\;\frac{3\pi\gamma\rho}{2\gamma_2 k}\; \int_0^{b_n} \; \left [
\frac{\pi}{2} -{\rm Si}\left (\frac{kr^2}{L}\right )\right ] r\; dr \; ,
$$
where
$$
{\rm Si}(x) \equiv \; \int_0^x\; \frac{\sin t}{t}\; dt = \frac{\pi}{2} +
\int_\infty^x\; \frac{\sin t}{t}\; dt
$$
is the integral sine. Introducing notation (48), we come to $g(\beta)$
given in Eq. (51). In the same way, for the coupling (38) we find
$$
g'(\beta)=-\; \frac{3\pi\gamma\rho L}{4\gamma_2 k^2} \;
\int_0^{2\beta}\; {\rm Ci}(x)\; dx\; ,
$$
where
$$
{\rm Ci}(x) \equiv \int_0^x\; \frac{\cos t}{t}\; dt
$$
is the integral cosine. The found expressions for the effective couplings can
also be transformed by means of the integrals
$$
\int\; {\rm Si}(x)\; dx = x{\rm Si}(x) +\cos x\; , \qquad
\int\; {\rm Ci}(x)\; dx = x{\rm Ci}(x) -\sin x\; .
$$
This yields
$$
g(\beta)=\; \frac{3\pi\gamma\rho L}{4\gamma_2k^2}\; \left [
\pi\beta -2\beta\;{\rm Si}(2\beta) + 1 -\cos(2\beta) \right ]\; ,
$$
$$
g'(\beta)=\; \frac{3\pi\gamma\rho L}{4\gamma_2k^2}\; \left [
\sin(2\beta)  -2\beta\;{\rm Ci}(2\beta)\right ]\; .
$$


\newpage

\begin{thebibliography}{99}

\bibitem{1}
M.C. Cross, P.C. Hohenberg, Rev. Mod. Phys. 65 (1993) 851.

\bibitem{2}
G. Dangelmayr, L. Kramer, in: Evolution of Spontaneous Structures
in Dissipative Continuous Systems, F.H. Busse, S.C. M\"uller (Eds.),
Springer, Berlin, 1998, p. 1.

\bibitem{3}
J.E. Mayer, M.G. Mayer, Statistical Mechanics, Wiley, New York, 1977.

\bibitem{4}
J. Rau, B. M\"uller, Phys. Rep. 272 (1996) 1.

\bibitem{5}
V.I. Yukalov, Physica A 234 (1997) 725.

\bibitem{6}
A.J. Lichtenberg, M.A. Liberman, Regular and Chaotic Dynamics,
Springer, New York, 1992.

\bibitem{7}
V.I. Yukalov, Phys. Rep. 208 (1991) 395.

\bibitem{8}
P. Coulett, L. Gil, F. Rocca, Opt. Commun. 73 (1989) 403.

\bibitem{9}
Y.S. Kivshar, B. Luther-Davies, Phys. Rep. 298 (1998) 81.

\bibitem{10}
K. Staliunas, Phys. Rev. A 48 (1993) 1573.


\bibitem{11}
L.A. Lugiato, Phys. Rep. 219 (1992) 293.

\bibitem{12}
L.A. Lugiato, F. Prati, L.M. Narducci, P. Ru, J.R. Tredicce, D.K. Bandi,
Phys. Rev. A 37 (1988) 3847.

\bibitem{13}
J.R. Tredicce, E.J. Quel, A.M. Ghazzawi, C. Green, M.A. Pernigo, 
L.M. Narducci, L.A. Lugiato, Phys. Rev. Lett. 62 (1989) 1274.

\bibitem{14}
L.A. Lugiato, G.L. Oppo, J.R. Tredicce, L.M. Narducci, M.A. Pernigo,
J. Opt. Soc. Am. B 7 (1990) 1019.

\bibitem{15}
C. Green, G.B. Mindlin, E.J. D'Angelo, H.G. Solari, J.R. Tredicce,
Phys. Rev. Lett. 65 (1990) 3124.

\bibitem{16}
M. Brambilla, F. Battipede, L.A. Lugiato, V. Penna, F. Prati, C. Tamm, 
C.O. Weiss, Phys. Rev. A 43 (1991) 5090.

\bibitem{17}
M. Brambilla, L.A. Lugiato, V. Penna, F. Prati, C. Tamm, C.O. Weiss,
Phys. Rev. A 43 (1991) 5114.

\bibitem{18}
F.T. Arecchi, G. Giacomelli, P.L. Ramazza, S. Residori, Phys. Rev. Lett.
65 (1990) 2531.

\bibitem{19}
F.T. Arecchi, G. Giacomelli, P.L. Ramazza, S. Residori, Phys. Rev. Lett.
67 (1991) 3749.

\bibitem{20}
F.T. Arecchi, Physica D 86 (1995) 297.

\bibitem{21}
D. Dangoisse, D. Hennequin, C. Lepers, E. Louvergneaux, P. Glorieux,
Phys. Rev. A 46 (1992) 5955.

\bibitem{22}
G. Huyet, J.R. Tredicce, Physica D 96 (1996) 209.

\bibitem{23}
G. Huyet, S. Rica, Physica D 96 (1996) 215.

\bibitem{24}
G. Nicolis, I. Prigogine, Self-Organization in Nonequilibrium
Systems, Wiley, New York, 1974.

\bibitem{25}
I. Pastor, J.M. Guerra, Appl. Phys. B 51 (1990) 342.

\bibitem{26}
I. Pastor, F. Encinas-Sanz, J.M. Guerra, Appl. Phys. B 52 (1991) 184.

\bibitem{27}
I. Pastor, V.M. P\'erez-Garcia, F. Encinas-Sanz, M.J. Guerra, L. Vazquez,
Physica D 66 (1993) 412.

\bibitem{28}
V.M. P\'erez-Garcia, J.M. Guerra, Phys. Rev. A 50 (1994) 1646.

\bibitem{29}
V.M. P\'erez-Garcia, I. Pastor, J.M. Guerra, Phys. Rev. A 52 (1995) 2392.

\bibitem{30}
F. Encinas-Sanz, J.M. Guerra, Opt. Lett. 21 (1996) 1153.

\bibitem{31}
F.A. Korolev, G.V. Abrosimov, A.I. Odintsov,  V.P. Yakunin, Opt.
Spectrosc. 28 (1970) 290.

\bibitem{32}
G.V. Abrosimov, Opt. Spectrosc. 31 (1971) 54.

\bibitem{33}
F.A. Korolev, G.V. Abrosimov, A.I. Odintsov, Opt. Spectrosc. 33 (1972) 399.

\bibitem{34}
V.I. Ishenko, V.N. Lisitsyn, A.M. Razhev, S.G. Rautian, A.M. Shalagin,
JETP Lett. 19 (1974) 346.

\bibitem{35}
F.A. Korolev, A.I. Odintsov, E.G. Turkin, V.P. Yakunin, Quant. Electron.
2 (1975) 413.

\bibitem{36}
V.I. Emelyanov, V.I. Yukalov, Opt. Spectrosc. 60 (1986) 385.

\bibitem{37}
V.I. Yukalov, J. Mod. Opt. 35 (1988) 35.

\bibitem{38}
V.I. Yukalov, J. Mod. Opt. 37 (1990) 1361.

\bibitem{39}
V.I. Yukalov, Laser Phys. 1 (1991) 81.

\bibitem{40}
A.V. Andreev, V.I. Emelyanov, Y.A. Ilinski, Cooperative Effects
in Optics, Institute of Physics, Bristol, 1993.

\bibitem{41}
V.I. Yukalov, Laser Phys. 1 (1991) 85.

\bibitem{42}
V.I. Yukalov, Laser Phys.  8 (1998) 1182.

\bibitem{43}
V.I. Yukalov, Phys. Rev. Lett. 75 (1995) 3000.

\bibitem{44}
V.I. Yukalov, Phys. Rev. B 53 (1996) 9232.

\bibitem{45}
N.N. Bogolubov, Y.A. Mitropolsky, Asymptotic Methods in the
Theory of Nonlinear Oscillations, Gordon and Breach, New York, 1961.

\end{thebibliography}
\end{document}